\documentclass[english,aps,prl,showpacs,amsmath,amssymb,twocolumn]{revtex4}
\usepackage{epsfig}
\usepackage{bm}
\usepackage{bbold}
\usepackage{latexsym}
\usepackage{color}

\begin{document}
	\title{On the uncertainty principle in Rindler and Friedmann spacetimes}
	\author{Thomas Sch\"urmann}
	\email{<t.schurmann@icloud.com>}
	\affiliation{D\"usseldorf, Germany}

\begin{abstract}
	We revise the extended uncertainty relations for the Rindler and Friedmann spacetimes recently discussed by Dabrowski and Wagner in \cite{DW19}. We reveal these results to be coordinate dependent expressions of the invariant uncertainty relations recently derived for general 3-dimensional spaces of constant curvature in \cite{TS18}. Moreover, we show that the non-zero minimum standard deviations of the momentum in \cite{DW19} are just artifacts caused by an unfavorable choice of coordinate systems which can be removed by standard arguments of geodesic completion. 
\end{abstract}

\maketitle

\section{I.\quad Introduction}
	One of the open problems in contemporary physics is the unification of quantum mechanics and general relativity in the framework of quantum gravity. Quantum gravity phenomenology studies quantum gravity effects in low-energy systems. The basis of such phenomenological models is the generalized or the extended uncertainty principle (GUP/EUP) \cite{KMM95}\cite{BC05}\cite{P08}\cite{M10}. The main characteristics of such investigations typically consist of modified commutation relations between position and momentum, including a linear or quadratic dependence on the position or the momenta, as well as certain phenomenological parameters to highlight the terms originating from the linear and quadratic contribution related to the scale at which quantum-gravitational effects are expected to become relevant \cite{LP17}. However, it should be mentioned that both the GUP and the EUP are originally derived in the literature by using modified commutation relations introduced {\it ad hoc}.

	In \cite{CBL16}, a translation operator acting in a space with a diagonal metric has been introduced to develop a derivation of the EUP from first principles. It has been shown that for any (sufficiently smooth) metric expanded up to the second order, this formalism naturally leads to an EUP and to a minimum non-zero standard deviation of the momentum. This gives reason to expect the existence of even higher order corrections in the EUP if the metric had been considered without approximation.  

	Rigorous mathematical derivations of uncertainty principles on Riemannian manifolds are hard to obtain. The problem already becomes apparent for quantum particles on the circle and on the sphere \cite{T04}. One of the difficulties for these systems is the position uncertainty measure for the particle (or the wave function spread measure). This is a consequence of the issue related to the choice of the operator for the azimuthal angle. This problem certainly holds for any closed manifold. For the 2-sphere the situation is even more complicated because of the absence of a self-adjoint momentum operator related to the azimuthal angle. This problem can be solved by the definition of a special coordinate system on the 2-sphere \cite{GP04}. 

Recently, it has been mentioned by  Dabrowski and Wagner \cite{DW19} that there are exact formulas for the EUP in the case of Rindler and Friedmann horizons and that these can be expanded to obtain asymptotic forms known from the previous literature.
The approach of \cite{DW19} requires a foliation of spacetime into hypersurfaces of constant time and so one considers the 3-dimensional spatial part of the corresponding spacetime metric. The underlying idea of the approach in \cite{DW19} is that the measurement of momentum depends on a given spacetime background recently introduced in \cite{TS18}\cite{TS09}. In order to measure the momentum one needs to consider a measure of position uncertainty. This is given by a domain $D$ (typically the geodesic ball $B_r$) with boundary $\partial D$ characterized by its geodesic radius  $r$ or diameter $d$ and Dirichlet boundary conditions such that the wave function of the particle is confined in $D$. The method then reduces to the solution of an eigenvalue problem for the wave function $\psi$
\begin{eqnarray}\label{EVP0}
\Delta \psi+\lambda \psi=0
\end{eqnarray}
inside $D$ with the requirement that $\psi=0$ on the boundary $\partial D$, while $\lambda$ denotes the eigenvalue and $\Delta$ is the Laplace-Beltrami operator of the corresponding manifold. Then, one can write the following general inequality \cite{TS18}  
\begin{eqnarray}\label{UP0}
\sigma_p\geq \hbar\, \sqrt{\lambda_1},
\end{eqnarray}
where $\lambda_1$ denotes the first Dirichlet eigenvalue of the problem. For the general class of 3-dimensional Riemannian manifolds of constant curvature $K$, there is a closed form solution and it was found that \cite{TS18} 
\begin{eqnarray}\label{UP1}
\sigma_p \,r\geq \pi\hbar\, \sqrt{1-\frac{K}{\pi^2}\,r^2},
\end{eqnarray}
where the corresponding position uncertainty of the particle is represented by the radius $r$ of the associated geodesic ball. It should be mentioned that this uncertainty relation is independent of the coordinate system (diffeomorphism invariance) and not of the same kind as the ordinary EUP or GUP because it features the characteristic length of the confinement corresponding to $r$. Thus, $r$ should rather be interpreted as uncertainty and does not describe the standard deviation of position \cite{TS18}\cite{TS09}. 
Both the Rindler geometry and the foliations of the Friedmann cosmology at a given instant of time are spaces of constant curvature $K$. For the Rindler space we have $K=0$ and from (\ref{UP1}), one simply obtains the uncertainty relation   
\begin{eqnarray}\label{UP11}
\sigma_p \,r\geq \pi\hbar,
\end{eqnarray}
for $0\leq r < \infty$. In the Friedmann spacetime, the sectional curvature of the spacelike hypersurface depends on the cosmological time $\tau$ and is given by 
\begin{eqnarray}\label{KFM}
K_\tau=\frac{k}{a^2(\tau)},
\end{eqnarray}
with $k=0,\pm 1$, corresponding to a flat, closed or open spatial geometry and $a(\tau)$ is the time-dependent scaling function associated with the Friedmann solution of Einstein's field equations. According to (\ref{UP1}), for a given instant of time, the uncertainty relation in the Friedmann cosmology is given by 
\begin{eqnarray}\label{UP12}
\sigma_p \,r\geq \pi\hbar\sqrt{1-\frac{k}{(\pi a)^2}\,r^2},
\end{eqnarray}
for $0\leq r \leq \pi a(\tau)$, if $k=1$, or $r\geq 0$, if $k=0$ or $-1$.
Although this closed form expression is valid for all three possibilities of $k$, we want to note that the corresponding physical context is very different.
For $k=1$, the space is isometric to the unit sphere and the ball with maximum position uncertainty is reached for $r\to \pi a(\tau)$, corresponding to the total domain of measure $2\pi^2 a^3$. In this case, the right-hand side of (\ref{UP12}) approaches zero such that the momentum dispersion can be arbitrary small although the position uncertainty is still finite.
For $k=-1$, the space is isometric to the unit ball with the standard metric induced by the Lorentzian 4-space.  
In this case a remarkable fact is given when the position uncertainty $r$ tends to infinity
while $a(\tau)$ is finite. Then, we obtain the global lower bound of $\sigma_p\geq\hbar/a(\tau)$. On the other hand, when $r$ is kept finite but $a(\tau)$ approaches zero, then the standard deviation of the momentum tends to infinity \cite{TS18}. 

The appeal of these inequalities is that they are independent of the coordinate system. Moreover, inequality (\ref{UP12}) is universally applicable to any kind of scaling function $a(\tau)$ which is a solution of the Friedmann equation (e.g. for matter, radiation, curvature or even mixtures of them as given by the Lambda-CDM model).
These scaling factors typically approach infinity for $\tau \to \infty$, which implies that (\ref{UP12}) approaches the inequality (\ref{UP11}) of flat Minkowski space. Nevertheless, the fact that the right-hand side of (\ref{UP1}) is determined by the spatial curvature $K$ of the foliation is a mathematically rigorous result and might be pathbreaking for other approaches. For example, in the ordinary EUP it is argued \cite{BC05}\cite{P08} that in an Anti-de Sitter background the Heisenberg principle should be modified by introducing the cosmological constant $\Lambda=-3/l_\text{H}^2$, with $l_\text{H}$ the Anti-de Sitter radius, as \cite{BC05}
\begin{eqnarray}\label{EUP}
\sigma_p \sigma_x\geq \frac{\hbar}{2}\, \left[1+\frac{\sigma_x^2}{l_\text{H}^2}\,\right],
\end{eqnarray}
where it is assumed by convention that $l_\text{H}^2<0$ for de Sitter spacetime, and $l_\text{H}^2>0$ for the Anti-de Sitter case. Now, the left-hand side of this inequality is founded in the spatial hypersurface at a given instant of (cosmic) time. On the other hand, it is well known that the spatial curvature of the de Sitter spacetime foliation approaches the flat Euclidean space for infinite cosmic times. At least in this limit, the right-hand side of (\ref{EUP}) should approach the bound $\hbar/2$ of the ordinary Heisenberg relation. However, this is not the case because $l_\text{H}$ is by definition a constant quantity. A discussion of the uncertainty principle (\ref{UP1}) applied to the de Sitter and Anti-de Sitter background has recently been given by the author in \cite{TS20}. 

By this argumentation it becomes clear that Dabrowski and Wagner's intention in \cite{DW19} is somewhat doubtful because the different meaning of curvatures applied in both approaches (\ref{UP1}) and (\ref{EUP}). 
The uncertainty relations of Dabrowski and Wagner in \cite{DW19} look very different from (\ref{UP11}) and (\ref{UP12}) and they are much more complicated. Obviously, this is because they are not written in an invariant representation, but are related to special coordinate systems.

In the following two sections, we will discuss the statements made in \cite{DW19}. The inequality corresponding to the Rindler vacuum can be derived from (\ref{UP11}) because both the spatial curvature of the 3-dimensional space and the curvature of the 4-dimensional Minkowski spacetime are equal and zero. For the Friedmann cosmology we show that the time-independent horizon specified in \cite{DW19} is not compatible with the spatial curvature $K_\tau$ in (\ref{KFM}) and (\ref{UP12}). We also show that the non-zero minimum standard deviations of the momentum stated in \cite{DW19} can be removed by standard arguments of geodesic completion. A comment is given at the end. 

\section{II.\quad The uncertainty principle \qquad\qquad\qquad\qquad in Rindler space}
The approach of \cite{DW19} requires a foliation of spacetime into hypersurfaces of constant time and so one considers only the spatial part of the Rindler metric which is of the form \cite{DW19}\cite{P10}
\begin{eqnarray}\label{R-metric}
ds^2=\frac{c^2}{2\alpha x}\, dx^2+dy^2+dz^2
\end{eqnarray}
with acceleration $\alpha$ describing a boost in the $x$-direction as applied to the Minkowski space, $c$ the speed of light, and $y$ and $z$ denoting the components of the metric perpendicular to the $x$ direction in Rindler space \cite{DW19}. Without loss of generality, we have chosen the boost of acceleration in the direction of $x$. Let
\begin{eqnarray}\label{gij}
(g_{ij})=\text{diag}\left(\frac{c^2}{2\alpha x},1,1\right)
\end{eqnarray}
be the corresponding 3-dimensional metric. For the following argumentation we briefly introduce the formal representation of a geodesic ball in Rindler space. A geodesic ball in Rindler space can be obtained by a suitable coordinate transformation to the Euclidean space:
\begin{eqnarray}\label{coord-trans}
X=\left(l_0 x\right)^\frac{1}{2},\quad Y=y,\quad Z=z,
\end{eqnarray}
for $x\geq 0$ and the abbreviation $l_0=2c^2/\alpha$.
The corresponding metric in the new (Euclidean) coordinates $X,Y,Z$ is simply given by
\begin{eqnarray}\label{dij}
(g_{ij}) \longrightarrow   (\delta_{ij})=\text{diag}\left(1,1,1\right).
\end{eqnarray}
Now, the boundary of the 3-sphere of radius $r$ centered around the position $(a,0,0)$ is just given by the algebraic expression 
\begin{eqnarray}\label{S3}
(X-a)^2+Y^2+Z^2=r^2,
\end{eqnarray}
The corresponding geodesic ball in Rindler coordinates $x,y,z,$ is obtained by substitution of (\ref{coord-trans}) to (\ref{S3}) and reads  
\begin{eqnarray}\label{Rindler_ball}
&&\left|\sqrt{l_0 x}-a\right|\leq\sqrt{r^2-y^2-z^2}\qquad\text{with}\\
\nonumber\\
&&y^2+z^2\leq r^2,\qquad 0\leq y,z\leq r. 
\end{eqnarray}

\begin{figure}[t]
	\begin{center}
		\psfig{file=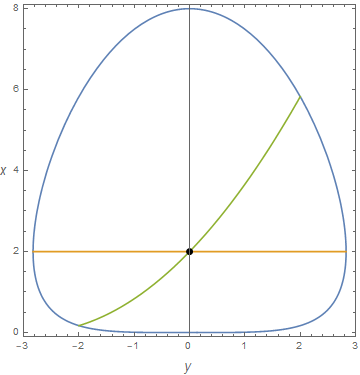,width=8.0cm, height=8.0cm}
		\parbox{8.0cm}{
			\caption{Projection of the 3-sphere in Rindler space (blue) onto the $x$-$y$-plane for $\alpha=1/2$ and $c=1$ ($l_0=4$). The acceleration is in $x$-direction, $y$ (and $z$) are perpendicular to the acceleration (see text). The black dot at $(l_0/2,0,0)$ is the center of the sphere in Rindler coordinates with radius $r=l_0/\sqrt{2}$. Also shown are diameters of geodesic length $2r$ (orange and green). The corresponding Euclidean 3-sphere of the same diameter is shown in Fig.\,\ref{fig2}.     
				\label{fig1}}}
	\end{center}
\end{figure}

\begin{figure}[t]
	\begin{center}
		\psfig{file=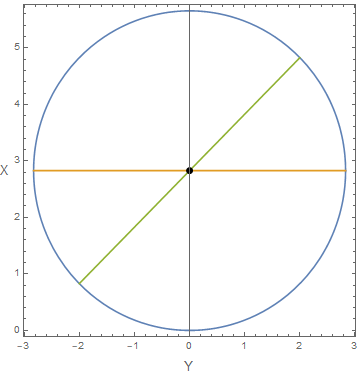,width=8.0cm, height=8.0cm}
		\parbox{8.0cm}{
			\caption{Projection of the 3-sphere in Euclidean space (blue) onto the $X$-$Y$-plane for $c=1$ ($l_0=4$). The black dot is at the center $(l_0/\sqrt{2},0,0)$ of the sphere with radius $r=l_0/\sqrt{2}$. Diameters of length $2r$ are also shown (green and orange), see also Fig.\,\ref{fig1}.     
				\label{fig2}}}
	\end{center}
\end{figure}

Rindler observers are accelerated
	with respect to inertial observers and additionally their acceleration differs
	in their positions (the closer to $x = 0$ they are, the larger is their
	acceleration). Because of the effect of Lorentz contraction, the endpoints
	(closer to $x = 0$) of a line of points must accelerate harder than the front points and this is
	reflected in deforming a geodesic ball.
Because of the axial symmetry with respect to the  $x$-direction, the corresponding 3-sphere can be properly expressed for $z=0$, see Fig.\,\ref{fig1} and \ref{fig2}. For instance, the vertical distance between the center (dot) and the north pole is different from the coordinate distance of $(0,0,0)$ to the center, although the geodesic distance of both is identical to $r$. 
So, if one wanted to express the position uncertainty relative to the $x$-direction (as has been done in \cite{DW19}), then one must take into account that the vertical coordinate distance is dependent on which position the circle is located in this direction. Actually, such a dependency is somewhat cumbersome and hard to handle by the observer. The appropriate choice of the position uncertainty should be the geodesic radius or diameter of the ball, which is constant and independent of its position in Rindler space. 
We briefly express the corresponding boundary value problem in Rindler coordinates. 
According to the metric (\ref{R-metric}), the Laplace-Beltrami operator of the problem is given by 
\begin{eqnarray}\label{LaplaceBeltrami}
\Delta =\frac{2\alpha}{c^2}\left( x\,\frac{\partial^2}{\partial x^2}+\frac{1}{2}\,\frac{\partial }{\partial x}\right)+\frac{\partial^2 }{\partial y^2}+\frac{\partial^2 }{\partial z^2},
\end{eqnarray}
so that the associated eigenvalue problem (\ref{EVP0}) is given by the following 3-dimensional partial differential equation 
\begin{eqnarray}\label{EVP}
\frac{2\alpha}{c^2}\left( x\,\frac{\partial^2\psi}{\partial x^2}+\frac{1}{2}\,\frac{\partial\psi}{\partial x}\right)+\frac{\partial^2 \psi}{\partial y^2}+\frac{\partial^2 \psi}{\partial z^2} + \lambda \psi=0.
\end{eqnarray}
Instead, to solve this equation in Rindler coordinates, as has been done in \cite{DW19}, here we follow an alternative approach by applying the coordinate transformation (\ref{coord-trans}) to equation (\ref{EVP}). After a few algebraic steps, we simply obtain
\begin{eqnarray}\label{EVP1}
\frac{\partial^2 f}{\partial X^2}+\frac{\partial^2 f}{\partial Y^2}+\frac{\partial^2 f}{\partial Z^2} + \lambda f=0,
\end{eqnarray}
while $f=f(X,Y,Z)\equiv\psi(x,y,z)$ is defined in the ordinary Euclidean space equipped with the standard metric (\ref{dij}) and the simple boundary condition 
\begin{eqnarray}\label{EVP2}
f(X,Y,Z)=0\qquad \text{for}\qquad (X,Y,Z)\in \partial S^3_r.
\end{eqnarray}
This problem has already been discussed in \cite{TS18} and the result is given by (\ref{UP1}), for $K=0$. In contrast to the result in \cite{DW19}, the acceleration does not explicitly occur in the invariant representation (\ref{UP11}).

More precisely, let us discuss the statement (18) of Dabrowski and Wagner in \cite{DW19}. Therein, it is proposed to express the position uncertainty by the coordinate $\Delta x$ in the direction of acceleration and the associated 1-dimensional domain of position uncertainty is taken to be the interval $I=[l_0-\Delta x,l_0+\Delta x]$, with $l_0=2c^2/\alpha$. At this point we want to mention that the corresponding (coordinate) distance $d_x:=2 \Delta x$ in this direction is not equal to the geodesic diameter which is $d=2\,r$. Our starting point to understand the inequality (18) of \cite{DW19} is to express the geodesic radius $r$ of (\ref{UP11}) in terms of $\Delta x$.
Therefore, we first define the north pole and the south pole $x_\pm=l_0\pm\Delta x$ of the ball in Rindler space. Applying the coordinate transformation (\ref{coord-trans}) to $x_\pm$, we find the dependency
\begin{eqnarray}\label{Dd0}
d_x=\frac{2 a}{l_0}\,d.
\end{eqnarray}
or equivalently  
\begin{eqnarray}\label{Dd1}
\Delta x=\frac{2 a}{l_0}\,r.
\end{eqnarray}
It follows that the position uncertainty $\Delta x$ does not only depend on the measure of $I$ but also on the position at which the measurement is performed in space. This is certainly an unfavorable property in the choice of $\Delta x$. However, let us apply (\ref{Dd1}) to express the inequality (\ref{UP11}) in terms of $\Delta x$. To eliminate the dependency on $a$, we consider the pre-image of $x_\pm$ under coordinate transformation (\ref{coord-trans}), that is 
\begin{eqnarray}\label{x}
x_\pm=\frac{1}{l_0}\,(a \pm r)^2
\end{eqnarray}
and by a few algebraic manipulations we get the equivalent expression 
\begin{eqnarray}\label{x1}
\frac{2a}{l_0}=\sqrt{1+\frac{\Delta x}{l_0}}+\sqrt{1-\frac{\Delta x}{l_0}}.
\end{eqnarray}
By substitution into (\ref{Dd1}), we obtain the geodesic radius $r$ in terms of $\Delta x$ and the acceleration $\alpha$, that is     
\begin{eqnarray}\label{Dd3}
r=\frac{\Delta x}{\sqrt{1+\frac{\Delta x}{l_0}}+\sqrt{1-\frac{\Delta x}{l_0}}}.
\end{eqnarray}
This expression is now applied for $r$ in the 1-dimensional version of (\ref{UP11}) given in \cite{TS09}. The square root terms are subsequently rearranged on the right-hand side, such that we find
\begin{eqnarray}\label{Dd4}
\sigma_p \Delta x\geq\frac{\pi \hbar}{2}\left(\sqrt{1+\frac{\Delta x}{l_0}}+\sqrt{1-\frac{\Delta x}{l_0}}\right).
\end{eqnarray}
We finally apply the binomial formula to the terms with square roots to get 
\begin{eqnarray}\label{Dd5}
\sigma_p \Delta x\geq \pi\hbar\,\frac{\frac{\Delta x}{l_0}}{\sqrt{1+\frac{\Delta x}{l_0}}-\sqrt{1-\frac{\Delta x}{l_0}}}.
\end{eqnarray}
By this derivation it becomes obvious that the complicated square root expression (\ref{Dd5}) is just a representation of the geodesic radius with respect to the coordinate dependent projection in the direction of acceleration. 
Since in the approach of \cite{DW19} the uncertainty of position is restricted by $\Delta x\leq l_0$, we see that the minimum possible $\sigma_p$ in the Rindler chart will be $\pi\hbar\sqrt{2}/l_0>0$. 
The meaning of the bound $\Delta x\leq l_0$ seems to be similar to the meaning of the Unruh temperature - both appear in Rindler frames for Rindler coordinates.  
However, the lower bound for $\sigma_p$ only holds for measurements which are performed for the set of balls with the center at $(l_0/2,0,0)$ in Rindler space (see Fig.\,\ref{fig1}). 
Alternatively, if we consider balls with the center at position $(r^2/l_0,0,0)$ in Rindler space, or equivalently $a=r$ in (\ref{Dd1}), then we have $r=(l_0\Delta x/2)^{1/2}$. By substitution into the 1-dimensional version of (\ref{UP11}), we obtain 
\begin{eqnarray}\label{Dd6}
\sigma_p \Delta x\geq \pi\hbar\,\sqrt{\frac{\Delta x}{2 l_0}}
\end{eqnarray}
and there is no restriction for $\Delta x$ in (\ref{Dd6}). As a consequence, the greatest lower bound of $\sigma_p$ is $0$, which can be obtained for $\Delta x\to\infty$. As we can see, the value of the greatest lower bound of $\sigma_p$ depends on the position at which the measurement process is performed in Rindler space. However, what we can be sure about is that $\sigma_p\to 0$ is possible.  

We already know from literature that the Rindler spacetime cannot be geodesically complete, because it covers only a portion of the original Minkowski spacetime, which is geodesically complete. However, the Rindler spacetime (and its 3-dimensional foliation) can be extended to the Minkowski spacetime (or the 3-dimensional Euclidean subspace) such that there is no longer any singularity in the metric components. 

\section{III.\quad The uncertainty principle in \qquad Friedmann cosmologies}

For the general case of the Friedmann cosmos let us proceed under the assumption that the universe is homogeneous and isotropic. Then there exists a  one-parameter family of spacelike hypersurfaces $\Sigma_\tau$, foliating the spacetime into pieces labelled by the proper time, $\tau$, of a clock carried by any isotropic observer. In these coordinates the spacetime metrics can be written as
\begin{eqnarray}\label{RW1}
ds^2=-c^2 d\tau^2+a^2(\tau)\left\{%
\begin{array}{ll}
d\chi^2+\sin^2\chi\,d\Omega^2, & 0\leq\chi\leq\pi\\
d\chi^2+\chi^2\,d\Omega^2,  & \chi\geq 0\\
d\chi^2+\sinh^2\chi\,d\Omega^2,  & \chi\geq 0 \\
\end{array}%
\right.\nonumber
\!\!\!\!\!\!\\
\end{eqnarray}
where the three possibilities beside the bracket correspond to the three possible spatial geometries, either the flat Euclidean space ($k=0$), the closed unit sphere ($k=1$) or the open unit hyperboloid ($k=-1$) \cite{Wald}. The general form of (\ref{RW1}) is called the Robertson-Walker cosmological model. The scale function $a(\tau)$ is given by the solution of the associated Friedmann equations. Without any further specification of the scale function, the corresponding time slices at any instant of $\tau$ are 3-dimensional subspaces $\Sigma_\tau$ of constant sectional curvature given by (\ref{KFM}), where the corresponding uncertainty relation has already been introduced in (\ref{UP12}). 

On the other hand, in section 4 of \cite{DW19}, Drabowski and Wagner consider an alternative (isometric) representation of the Robertson-Walker form corresponding to
\begin{eqnarray}\label{FM}
ds^2=-c^2 d\tau^2+\frac{d\tilde r^2}{1-\frac{\tilde r^2}{\tilde r_\text{\tiny H}^2}} + \tilde r^2 d\Omega,
\end{eqnarray}
with the horizon $\tilde r_\text{\tiny H}$ defined by 
\begin{eqnarray}\label{hDW}
\tilde r_\text{\tiny H}^2=\frac{c^2}{H^2+\frac{k c^2}{a^2}}.
\end{eqnarray}
Here, we have slightly adjusted the notation in (\ref{FM}) by writing $\tilde r$ instead of $r$. The reason is that $\tilde r$ in the representation of (\ref{FM}) is not a geodesic coordinate and has to be distinguished from the geodesic radius $r$ defined in the previous sections. Moreover, in (\ref{hDW}) we used the notation $k$ instead of $K$ for the curvature index, while the latter has been applied in (22) of \cite{DW19}. Now, using the definition of the Hubble function $H=\dot{a}/a$, and after a few algebraic manipulations we see that expression (\ref{hDW}) is just equivalent to the Friedmann equation with curvature $k$ and cosmological constant $\Lambda=3/\tilde r_\text{\tiny H}^2$, given by
\begin{eqnarray}\label{FMvac}
\frac{\dot a^2}{c^2}=\frac{\Lambda}{3}\, a^2-k.
\end{eqnarray}
	Actually, this case corresponds to the standard de Sitter and Anti-de Sitter spacetime. At this point we see that $\tilde r_\text{\tiny H}$ is in fact not the curvature radius of the foliation in a given time slice and is therefore not appropriate to be used for the uncertainty relation under consideration (see \cite{TS20}). Instead, the correct curvature radius is corresponding to the scale factor $a(\tau)$. For reasons of comparison with \cite{DW19}, in the following we will define the notation (without tilde) given by
\begin{eqnarray}\label{red}
r_\text{\tiny H}\equiv\,a(\tau), 
\end{eqnarray}
instead of $\tilde r_\text{\tiny H}$ given in (\ref{hDW}).
With this notation, the approach in \cite{DW19} is to solve the associated Dirichlet boundary value problem in $\tilde r$ and to obtain the expression (29) in \cite{DW19}. As already mentioned in the introduction, this physical situation has already been treated by inequality (\ref{UP12}), with $k=1$. 
To compare our result with the statement (29) in \cite{DW19}, we first rewrite (\ref{UP12}) as follows
\begin{eqnarray}\label{UP13}
\sigma_p\geq \frac{\hbar}{r_\text{\tiny H}}\, \sqrt{\left(\frac{\pi}{r/r_\text{\tiny H}}\right)^2-1}.
\end{eqnarray}
We remember that the standard representation of the spatial part in (\ref{FM}) is based on the coordinate transformation 
\begin{eqnarray}\label{geo}
\tilde r=r_\text{\tiny H}\sin\left(\frac{r}{r_\text{\tiny H}}\right),
\end{eqnarray}
for $0\leq r< \pi r_\text{\tiny H}$, which is a relation between the geodesic radius $r$ and the coordinate $\tilde r$. We also keep in mind that the foliation is not completely covered for $\tilde r\to r_\text{\tiny H}$. Actually, this limit is only related to the "upper" hemisphere of $S^3_{r_\text{\tiny H}}$ corresponding to the polar angle of $\pi/2$ in spherical coordinates (see \cite{Wald}, p.\,116). For covering the complete space, one must also regard the lower hemisphere, which is not a priori contained in the representation (\ref{FM}). This fact will be taken into account by writing (\ref{geo}) in terms of the two branches
\begin{eqnarray}\label{bran}
-\frac{r}{r_\text{\tiny H}}&=&\pm\arccos\left(\frac{\tilde r}{r_\text{\tiny H}}\right)-\frac{\pi}{2}\nonumber\\
&=&2\arctan\left( f_\pm(\tilde r)\right) -\frac{\pi}{2}       
\end{eqnarray}
with
\begin{eqnarray}\label{bran1}
f_\pm(\tilde r)= \pm\tan\left(\frac{1}{2}\arccos\left(\frac{\tilde r}{r_\text{\tiny H}}\right)\right). 
\end{eqnarray}
\begin{figure}[b]
	\begin{center}
		\psfig{file=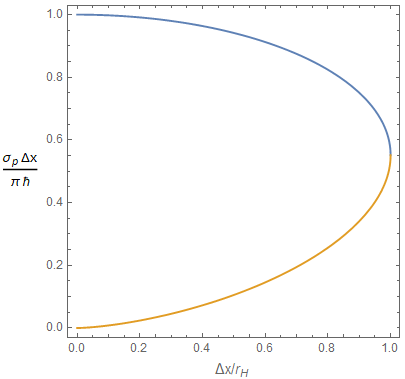,width=8.0cm, height=8.0cm}
		\parbox{8.0cm}{
			\caption{The uncertainty relation (\ref{UP12}) for the Friedmann cosmology in terms of the rescaled position uncertainty $\Delta x/r_\text{\tiny H}$ in units of $\pi\hbar$. In these units the uncertainty approaches its minimum value of 0. The curve is passed through the two branches (blue and orange) clockwise from top to bottom. For reasons of comparison, we have used the same notation as in Fig.\,2 of \cite{DW19}, that is $\tilde r\equiv\Delta x$  (see text). 
				\label{fig3}}}
	\end{center}
\end{figure}
\\
We reformulate this expression by using the half-angle formula
\begin{eqnarray}\label{ha}
\tan\frac{\xi}{2}=\sqrt{\frac{1-\cos \xi}{1+\cos \xi}},
\end{eqnarray}
for $\xi\in[0,\pi)$ and $\cos\xi:=\tilde r/r_\text{\tiny H}$, to obtain 
\begin{eqnarray}\label{bran2}
f_\pm(\tilde r)= \pm  \sqrt{\frac{1-\tilde r/r_\text{\tiny H}}{1+\tilde r/r_\text{\tiny H}}}.
\end{eqnarray}
By substitution of (\ref{bran}) into the denominator under the square root in (\ref{UP13}), we obtain the final result 
\begin{eqnarray}\label{UP14}
\sigma_p\,\tilde r\geq \hbar\,\frac{\tilde r}{r_\text{\tiny H}}\, \sqrt{\left(\frac{\pi}{2\arctan\left( f_\pm(\tilde r)\right) -\pi/2}\right)^2-1},
\end{eqnarray}
which reproduces expression (29) of \cite{DW19}, except that there are two signs in our result. 

In Fig.\,\ref{fig3}, we see the (total) uncertainty in terms of the notation applied in \cite{DW19}, that is $\tilde r\equiv \Delta x$. In Fig.\,2 of \cite{DW19}, there is only the upper (blue) part of the curve but not the lower branch (orange) and it is argued that the uncertainty approaches a minimum value of $\sqrt{3}/\pi$, for $\Delta x\to r_\text{\tiny H}$. As already mentioned above, this argumentation is incomplete, because it ignores the "lower" hemisphere of $S^3_{r_\text{\tiny H}}$. 
Instead, we consider the complete space and find the true minimum value of the uncertainty approaches zero (orange curve) when the space is completely covered by the position uncertainty such as $r\to \pi r_\text{\tiny H}\equiv \pi a(\tau)$. 
\section{IV.\quad Comment}  
As we know from the history of Riemannian geometry and general relativity, the property of diffeomorphism invariance is one of the most important features for the generalization of physical laws to curved spaces. For uncertainty principles given in 3-dimensional space this means that the applied measures of uncertainty should be chosen with caution. When the standard deviation of the momentum is based on the Laplace-Beltrami operator, then one can be sure that invariance under change of coordinates is satisfied. On the other hand, an obvious choice for the position uncertainty is hard to obtain if one is only concerned to apply the concept of standard deviation. As we have seen in the present contribution, fortunately the choice of a standard deviation in position space is not really necessary or even appropriate. Especially from the concept of projection-valued measures it becomes obvious, alternatively, to consider suitable spatial domains for the representation of position uncertainty. Moreover, from the theory of spectral analysis, we know that geodesic balls play an important role because these are the distinguished domains in many variational approaches. Since geodesic balls are uniquely classified  by their geodesic radius (or diameter) it becomes obvious that the geodesic radius is the appropriate measure for the representation of position uncertainty in curved spaces. For that reason it becomes clear why the requirement of coordinate invariance is hard to obtain by the standard GUP and EUP in literature. This fact makes their analysis and interpretation sometimes difficult. 
\\
\\
\\
\\

\end{document}